
\magnification=1200
\hsize=6truein
\def\p{\partial}
\def\pz{\p_z}
\def\pa{\p_{-h+1,1}}
\def\ta{t_{-h+1,1}}
\rightheadtext{additional symmetries, Grassmannian, and string equation}
\topmatter
\NoBlackBoxes
\title  Additional symmetries of KP, Grassmannian,\\
 and the string equation \endtitle
\author  L.A.Dickey  \endauthor
\affil University of Oklahoma, Norman, OK 73019  \endaffil
\address e-mail: ldickey\@nsfuvax.math.uoknor.edu \endaddress
\date April, 1992 \enddate
\endtopmatter
\document
\subheading{1. Introduction} The so-called string equation received the most
serious study during the last two or three years. It occured to be relevant
e.g. to the non-perturbative theory of 2-d quantum gravity \cite{1}. This is
the equation $[Q,P]=1$ where $Q$ and $P$ are
differential operators of the orders, say, $h$ and $p$. (The simplest of these
equations coincides with the Painlev\'e transcendent I). The string equation
resembles the old problem: to find two commuting operators $[Q,P]=0$ which is
tightly related to the theory of the Korteweg - de Vries (KdV) hieararchy of
equations. Therefore it is not surprising that one tries to construct a theory
for the string equation using the rich experience with the KdV or, more
general, with the Kadomtsev-Petviashvili (KP) hierarachy.

In a series of articles by Fukuma, Kawai and Nakayama \cite{2} and Kac and
Schwarz \cite{3} the solutions to this equation are related to elements of
Sato's
Grassmannian. More precisely, if, in addition to a condition that an element
$V$ of the Grassmannian ($V$ is a subspace of the space of functions $f(z)$ on
the unit circle) satisfies the equation $z^hV\subset V$ which guarantees that
the KP hierarchy reduces to KdV and the operator $Q$ is differential of the
order $h$, $V$ is also invariant under the action of the operator
$A=z^{-h+1}\pz
+(1-h)z^{-h}/2$ then this subspace determines a solution to the string
equation. Besides, it was established (\cite{4},\cite{2},\cite{3}) that this
condition is equivalent to some "Virasoro constraints" on the corresponding
$\tau$-function.

On the other hand, in the works by La \cite{5} and, virtually, by Goeree \cite
{6} (though
he does not say this explicitly) the problem is connected to the so-called
additional symmetries of the KdV hierarchies i.e. symmetries involving explicit
dependence of the independent variables. Both of them considered the case of
the lowest orders, $h=2,3$. The string equation is equivalent to the
requirement
that the operator does not depend on the parameter of an additional symmetry.

A problem arises. It is well-known that KP or KdV flows are represented by very
simple flows on Grassmannians. In particular, the operator $V\mapsto z^hV$ is
the infinitesimal operator of the $h$th KP flow. Thus, the invariance of an
element $V\in$Gr with respect to this operator implies independence of the
constructed operator $Q$ of the $h$th time variable and yields the reduction of
KP to the $h$th KdV (which means that the operator $Q$ is differential). Maybe,
the above operator $A$ plays the same role as $z^h$ but now for the group of
additional symmetries? The answer is yes, this is just the case, and our goal
is to show this.

Thus, we, first, find the relation between two lines of study: the works
concerned with Grassmannians and those based on additional symmetries. Second,
even if one is not dealing with the string equation, there exist groups of
additional symmetries and it is just interesting to find how they are
represented on the Grassmannian. We have not heard about works studying
this very natural problem. Third, in contrast to \cite{5} and \cite{6} we
consider arbitrary $h$. We use the techniques developed by Orlov and Shulman
\cite{7} (see also our book \cite{8}, sect.7.8). Finally, the proof of the
Virasoro constraints is straightforward in these frameworks and do not require
a "fermionisation" i.e. a modeling of the Grassmannian in the Fock
representation of a Clifford algebra as it was the case in \cite{2} and
\cite{3}. The obtained constraints are a little bit more general than those
written in the cited papers.
\vskip1\baselineskip
\subheading{2. Additional symmetries and the string equation} Let $L$ be a
pseudodifferential operator ($\Psi DO$) of the first order
$$ L=\p+u_1\p^{-1}+u_2\p^{-2}+...,~~\p=\p/\p x.$$ Let $L$ depend on some time
parameters $t_1,t_2,...$; $t_1=x$. The KP hierarchy is the set of equations
\cite{9}
$$ \p_nL=[L_+^n,L],~~\p_n=\p/\p t_n.$$ Here the subscript $+$ means choosing
only non-negative powers of $\p$. It is well-known that all the equations
commute and therefore they are symmetries to each other.

If for some positive integer $h$ the operator $L^h=Q$ is purely differential
then the KP hierarchy reduces to a generalized KdV hierarchy
$$ \p_nQ=[Q_+^{n/h},Q]$$ first introduced in [10].

Let $L$ be represented in the "formal dressing" form:
$$ L=\Phi\p\Phi^{-1},~~~\Phi =1+\sum_0^{\infty}w_i\p^{-i-1}.$$
Notice that $\Phi$ is determined up to the following gauge transformation:
multiplication on the right on arbitrary series $1+\sum_0^{\infty}c_i\p^{-i-1}$
with constant $c_i$.

Applying the operator $\Phi$ to $\exp\xi$ where $\xi=\sum_1^{\infty}t_kz^k$ we
obtain the Baker function $w(t,z)=\hat{w}(t,z)\exp\xi$, $$\hat{w}(t,z)=
1+\sum_0^{\infty}w_iz^{-i-1}.$$ The gauge transformation in terms of the Baker
function is the multiplication by a constant series $1+\sum_0^{\infty} c_iz^
{-i-1}$.

The Baker function satisfies the equations $$Lw=zw,~~\p_nw=L_+^nw.$$ In terms
of the dressing operator $\Phi$
the KP hierarchy is equivalent to $\p_n\Phi=-L_-^n\Phi$. The equations of the
hierarchy which can be written as $[\p_n-L_+^n,L]=0$ can be considered as
dressing of the obvious relation $[\p_n-\p^n,\p]=0$.

There exists, however, another operator commuting with $\p_n-\p^n$. This is
$\Gamma=\sum_1^{\infty}rt_r\p^{r-1}$. Let us dress this operator: $$M
=\Phi\sum_1^{\infty}rt_r\p^{r-1}\Phi^{-1}.\eqno{(1)}$$ Now,
$$ [\p_n-\p^n,\Gamma]=0 \Rightarrow [\p_n-L_+^n,M]=0$$ hence $[\p_n-L_+^n,M^k
L^m]=0$ for any $k,m\in \bold Z,~k\geq 0$. For each pair $k,m$ choose a
variable $t_{m,k}$ and write the equation
$$\p_{mk}L=-[(M^kL^m)_-,L]$$ which yields $$\p_{mk}L^h=-[(M^kL^m)_-,L^h].\eqno
{(2)}$$ It is easy to prove that $\p_{mk}$ commute with $\p_n$ (see, e.g. in
\cite{8}, Sect.7.8). This means that symmetries of the KP hierarchy are
constructed (they do not commute between themselves). These are "additional
symmetries" \cite{12}, \cite{7}.

Further we speak only on a special case of the symmetry, $k=1$ and $m=-h+1$
which is relevant to the theory of the string equation. Eq(2) becomes
$$\p_{-h+1,1}L^h=-[h^{-1}(ML^{-h+1})_-,L^h].\eqno{(3)}$$
(We rescaled this equation dividing the right hand side by $h$).
Let us consider a reduction of the KP hierarchy to the $h$th KdV i.e.
$L^h$ is purely differential and $L_-^h$ will be identified with zero. We have
$[M,L]=\Phi[\sum rt_r\p^{r-1},\p]\Phi^{-1}=-1$, $[M,L^h]=-hL^{h-1}$, and $$[ML^
{-h+1},L^h]=-h. \eqno{(4)}$$  Eqs (3) and (4) imply
$$ \p_{-h+1,1}L^h=[h^{-1}(ML^{-h+1})_+,L^h]+1. \eqno{(5)}$$

The operator $P=h^{-1}(ML^{-h+1})_+$ is of an infinite order. If we wish to
have
operators of a finite order, say $p$, we must take in $\Gamma$ only finite
number of variables distinct from zero, namely, $t_i$ with $i\leq p+h-1$.
The flows $\p_i$ do not preserve the order of the operator $P$ if $i>p+h-1$.

A question arises: why, instead of (3), one cannot take
$$\p_{-h+1,1}L^h=-{1\over h}[(L^{-h+1}M)_-,L^h].$$ Is this a new symmetry? No,
this is the same: $[L^{-h+1},M]_-=(-h+1)L^{-h}$ and $[[L^{-h+1},M]_-,L^h]=0$.
The same symmetry can also be written e.g. in the form
$$\p_{-h+1,1}L^h=-{1\over 2h}[(ML^{-h+1})_-+(L^{-h+1}M)_-,L^h]\eqno{(6)}$$
and we prefer this form in what follows.

Further, we can argue in the following way. Let us compare the well-known
problem of pairs of commuting operators, $[P,Q]=0$ with our equation $[P,Q]=1$.
In the first case we are looking for $L$ not depending on two time variables,
$t_h$ and $t_p$, then $L^h$ and $L^p$ turn out to be differential operators,
and $0=\p_pL^h=[L^p,L^h]$. The second problem is absolutely similar: the
following holds by virtue of (5):
\proclaim{Proposition} If the operator $L$ does not depend on the parameter
$t_h$ and on the additional symmetry variable
$t_{-h+1,1}$ then $L^h$ will be a differential operator and $[L^h,h^{-1}(ML^{
-h+1})_+]=1$ which is precisely the string equation $[P,Q]=1$. \endproclaim
\proclaim{Remark} Operator $M$ along with the string equation are not invariant
under the above gauge transformation of $\Phi$.\endproclaim
\vskip1\baselineskip
\subheading{3. $z$-operators and the Grassmannian} Now we shall discuss
the following problem. As it is well-known, the KP
flows are represented on the Grassmannian by very simple flows. What kind of
flows on the Grassmannian correspond to our new, additional symmetries?
We use the version of Sato's Grassmannian from Mulase's article \cite{12}.
The advantage of this approach is the more direct expression of the operator
$L$ in terms of the elements of the Grassmannian. In this subsection we recall
this construction (with slight modifications; that forces us to expose this
theory here).

The expressions
$$ G=G(\pz,z)=\sum_{j\geq 0,i\leq i_0}a_{ij}z^i\pz^j,~~\pz=\p/\p z$$
will be called $z$-operators in a standard form. (More general form: the sum of
any monomials in $z$ and $\pz$ which always can be reduced to the standard form
using commutation rules). They act in the space $H$ of formal expansions $f(z)=
\sum_{-\infty}^{\infty}f_kz^k$ on the unit circle (we do not discuss their
convergence, this theory is mostly formal algebraic; Segal and Wilson \cite{13}
gave
it more analytical character). Sometimes the coefficients $a_{ij}$ will depend
on additional parameters $t=(t_1,t_2,...)$. Operators $$G=1+\sum_{j\geq0,i<0}a_
{ij}z^i\pz^j$$ are called monic.

One can assign a $\Psi DO$ to any $z$-operator:
$$^TG(x,\p)=\sum a_{ij}x^j\p^i.$$ The left superscript $T$ means inversion:
we replace $z\mapsto \p,\pz\mapsto x$ and write the factors in an inverse
order. (The same rule holds in the above more general, not standard, case). It
is easy to see that: $$f\cdot g\mapsto{}^Tg\cdot{}^Tf$$ which is the corollary
of the commutation rule $$[z,\pz]=[x,\p]=-1$$ and means that the correspondence
between $\Psi DO$ and $z$-operators is an anti-isomorphism. Besides, a relation
$$G(\pz,z)\exp(xz)={}^TG(x,\p)\exp(xz) \eqno{(7)}$$ holds. If $G(\pz,z)$ is in
the standard form then these expressions are
$G(x,z)\exp(xz)\mathbreak ={}^TG(x,z)\exp(xz)$.

In what follows, we shall drop the left superscript $T$ for simplicity.
However, {\it we must remember that passing from $\Psi DO$ to $z$-operators or
back we must write all factors in inverse order}.

The Grassmanniann Gr consists of linear subspaces $V\subset H$ such that
the natural projection $V\rightarrow H_+$ is a one-to-one correspondence.

It is not difficult to show \cite{12}, \cite{3}:

\proclaim {Proposition 1} If $V\in$Gr is an element of a Grassmannian then
there is a monic $z$-operator $G$ such that $V=GH_+$.\endproclaim

\proclaim {Proposition 2} If a $z$-operator preserves $H_+$ then it must
involve only non-negative powers of $z$. \endproclaim

Let $t^*=(t_2,t_3,...)$ and $\xi^*=\xi(t^*,z)=\sum_2^{\infty}t_kz^k$. If $V$ is
an element of the Grassmannian then, generically, $V\exp(-\xi(t^*,z))$ is
another element of that. Then there is a monic $z$-operator $\Phi(t^*,\pz,z)$
such that\footnote{This equation differs from the similar one in \cite{3}.}
 $$ \exp(-\xi^*)V=\Phi(t^*,\pz,z)H_+.\eqno{(8)}$$ We assume the $z$-operator $
\Phi$ written in the standard form. Then $$\Phi(t^*,\pz,z)\exp(xz)=\Phi(t^*,x,
\p)\exp(xz)=\Phi(t^*,x,z)\exp(xz)$$ (see (7)). The function of $z$: $\exp(xz)$
(where $x$ is a parameter) belongs to $H_+$, therefore, $\Phi(t^*,x,z)\exp(xz)=
\Phi(t^*,\pz,z)\exp(xz)\in V\exp(-\xi(t^*,z))$ and \linebreak $\Phi(t^*,x,z)
\exp\xi(t,z)\in V$, here $t=t_1,t_2,...$ and $t_1=x$; $\xi(t,z)=\xi(t^*,z)+xz=
\sum_1^\infty t_iz^i$. According to the Sato-Segal-Wilson
theory this means that $$w_V(t,z)=\hat{w}_V\exp\xi(t,z)=\Phi(t^*,x,z)\exp\xi
(t,z)\eqno{(9)}$$ is the Baker function of the KP hierarchy for the operator
$$L=\Phi(t^*,x,\p)\p\Phi^{-1}(t^*,x,\p)\eqno{(10)}$$ related to $V\in$Gr.
Thus, we have
\proclaim{Theorem} If $V\in$Gr and $\Phi$ is defined by Eq(8) then $L$
defined by Eq(10) is a solution to KP.\endproclaim
Now we give another proof of this statement, more in the spirit of \cite{3}.
We start with the same relation $\Phi^{-1}(t^*,\pz,z)\exp(-\xi^*)V=H_
+$. Let $v\in V$ be a fixed element. Then $\Phi^{-1}\exp(-\xi^*)\cdot v=h_+\in
H_+$.Apply the operator $\p_k$ to this equality: $$-\Phi^{-1}\p_k\Phi\cdot\Phi^
{-1}\exp(-\xi^*)\cdot v-\Phi^{-1}z^k\exp(-\xi^*)\Phi^{-1}=\p_kh_+.$$
Put $$L(t^*,\pz,z)=\Phi^{-1}(t^*,\pz,z)z\Phi(t^*,\pz,z).$$ We have
$$-\Phi^{-1}\p_k\Phi\cdot\Phi^{-1}\exp(-\xi^*)\cdot v-L^k\Phi^{-1}\exp(-\xi^*)
\cdot v=\p_kh_+.$$ This can be written as $$(\Phi^{-1}\p_k\Phi+L_-^k)h_+=-L_+^k
h_+-\p_kh_+.$$
The right hand side belongs to $H_+$. Therefore $(\Phi^{-1}\p_k\Phi+L_-^k)
h_+\in H_+$. The element $h_+$ is an arbitrary element of $H_+$. Proposition 2
implies that the operator $\Phi^{-1}\p_k\Phi+L_-^k$ cannot contain negative
powers. On the other hand it does not contain non-negative powers, therefore it
vanishes and $$\p_k\Phi=-\Phi L_-^k=-\Phi(\Phi^{-1}z^k\Phi)_-.$$ Passing to the
$\Psi DO$ we have
$$\p_k\Phi(t^*,x,\p)=-(\Phi\p^k\Phi^{-1})_-\Phi. \eqno{(11)}
$$ If we let $L(t)=\Phi(t^*,x,\p)\p\Phi^{-1}(t^*,x,\p)$ then Eq(11)
is equivalent to the KP hierarchy for $L$.

If we do not discuss the anti-isomorphism between $\Psi DO$ and $z$-operators
and deal solely with $\Psi DO$ (where $t_1=x$ must not be treated separately
from other $t_k$) then instead of $t^*,x$ we can write just $t$.
\vskip1\baselineskip
\subheading{4. Additional symmetries and the Grassmannian}  Now, we can return
to the additional symmetry flow. This flow on the
Grassmannian is given by some operator: $V\rightarrow G(t)V$. The operator $G$
is determined up to a multiplication on the right by a factor preserving $V$.
Therefore, the infinitesimal operator acting on $V$ is determined up to a
summand preserving $V$ i.e. it can be considered as an operator $V\rightarrow
H/V$. (We are interested not in the mapping of vectors but in the mapping of
subspaces, the elements of the Grassmannian).

The equation (3) is, in terms of the dressing operator $\Phi(t^*,x,\p)$, $$\pa
\Phi=-h^{-1}(ML^{-h+1})_-\Phi.$$ However, with the same reason
we can take
$$\pa\Phi=-(2h)^{-1}(ML^{-h+1}+L^{-h+1}M)_-\Phi\eqno{(12)}$$ corresponding to
Eq(6) (the r.h.s. of two equations differ by a term
proportional to $L^{-h}\Phi=\Phi\p^{-h}$. This term is insignificant
because $\Phi$ is determined to within the right multiplication by
constant series $1+O(z^{-1})$). The form of equation (12) is the most
convenient
to obtain Virasoro constraints (see below).

In terms of the $z$-operators this is (inversion of the order!)
$$\pa\Phi(t^*,\pz,z)$$ $$=-(2h)^{-1}\Phi(L^{-h+1}M+ML^{-h+1})_-.$$
Let $v\in V$ be an element. Then there is a $h_+\in H_+$ such that $\exp(-\xi^*
)v =\Phi h_+$ (since $\exp(-\xi^*)V=\Phi H_+$).
We have $$\pa(\exp(-\xi^*)v)=\pa(\Phi h_+)$$ $$=-\Phi(2h)^{-1}(ML^{-h+1}+L^{-h+
1}M)_-\Phi^{-1}\exp(-\xi^*)v+\Phi\p_{-h+1,1}h_+.$$
Using the transformation $(~)_-=(~)-(~)_+$ and $\sum_2^{\infty}t_kkz^{k-1}\exp
(-\xi^*)=\mathbreak -\pz\exp(-\xi^*)$ we get $$\pa(\exp(-\xi^*)v)$$
$$=-(2h)^{-1}((\pz+\sum_2^{\infty}t_kkz^{k-1})z^{-h+1}+z^{-h+1}(\pz+\sum_2^
{\infty}t_kkz^{k-1}))\exp(-\xi^*)v$$ $$+(2h)^{-1}\Phi(ML^{-h+1}+L^{-h+1}M)_+h_+
+\Phi\p_{-h+1,1}h_+$$ $$=-(2h)^{-1}(\pz z^{-h+1}+z^{-h+1}\pz)(v\exp(-\xi^*))+
(2h)^{-1}v(2z^{-h+1}\pz)\exp(-\xi^*)$$ $$+\Phi\p_{-h+1,1}h_+
=-(2h)^{-1}\{2(z^{-h+1}\pz+(-h+1)z^{-h})v\}\cdot\exp(-\xi^*)$$ $$+(2h)^{-1}\Phi
(ML^{-h+1}+L^{-h+1}M)_+h_++\Phi\p_{-h+1,1}h_+.$$
 The second and the third term belong to $\Phi H_+=V\exp(-\xi^*)$ i.e.
their sum can be written as $v_1\exp(-\xi^*)$ where $v_1\in V$. The l.h.s.
of the equation is $\exp(-\xi^*)\pa v$ since the flow $\pa$ commutes with
$\p_k$ and, therefore, parameters $\ta$ and $t_k$ are independent. Dividing by
$\exp(-\xi^*)$ we obtain $$\p_{-h+1,1}v=-(2h)^{-1}(z^{-h+1}\pz+{-h+1\over 2}z^{
-h})v+v_1.$$ The last term does not play any role because we are interested in
operator \linebreak $\p_{-h+1,1}:~V\rightarrow H/V$. Thus, we proved

\proclaim{Proposition} The infinitesimal operator acting on an element $V$
of the Grassmannian which corresponds to the additional symmetry $1,-h+1$ is
$$A=-h^{-1}(z^{-h+1}\pz+{-h+1\over 2}z^{-h}).$$   \endproclaim
If an element $V$ of the Grassmannian is invariant under operators $z^h$ and
$A$ then: i) $L^h$ is a differential operator, ii) The element of the
Grassmannian $V$ is invariant under the flow in $\ta$. Then the monic operator
$\Phi$ does not depend on $\ta$, neither does the operator
$L^h$, and Eq(5) reduces to the string equation.
\vskip1\baselineskip
\subheading{5. Virasoro constraints} As we have seen, the string equation is
equivalent to  $\pa\Phi=0$. We shall
express this equation in terms of $\tau$-functions and obtain the so-called
Virasoro constraints. First we prove the formula: $$\pa\hat{w}(t,z)={1\over h}
(-z^{-h+1}\pz+{h-1\over 2}z^{-h}-\sum_1^{h-1}kt_kz^{k-h}+\sum_{h+1}^{\infty}kt_
k\p_{k-h})\hat{w}(t,z)\eqno{(13)}$$ where $\hat{w}(t,z)=\Phi(t^*,x,z)=(\Phi(t
^*,\pz,z)\exp(xz))\exp(-xz)$ is the coefficient in front of the exponent in the
Baker function (see (9)).

Let us start with
$$\pa\Phi(t^*,\pz,z)=-{1\over 2h}\Phi(ML^{-h+1}+L^{-h+1}M)_-$$ $$=-{1\over 2h}
\Phi\{\Phi^{-1}[(\pz+\sum_2^{\infty}t_kkz^{k-1})z^{-h+1}+z^{-h+1}(\pz+\sum_2^
{\infty}t_kkz^{k-1})]\Phi\}_-$$ $$=-{1\over 2h}[(-h+1)z^{-h}+2z^{-h+1}\pz+2\sum
_2^{h-1}kt_kz^{k-h}]\Phi$$ $$-{1\over 2h}\Phi[\Phi^{-1}2\sum_{h+1}^{\infty}kt_k
z^{k-h}\Phi]_- .$$(In the first term we skipped the subscript minus since
everything in the brackets is $O(z^{-1}))$. In the last term we write $\Phi[
\Phi^{-1}z^{k-h}\Phi]_-=-\p_{k-h}\Phi$.
Apply both the sides of the obtained operator equality to the element $\exp(xz)
\in H_+$: $$\pa\Phi(t^*,x,z)\exp(xz)$$ $$={1\over h}\{[{h-1\over 2}z^{-h}-z^{-
h+1}\pz]\Phi(t^*,x,z)\}\exp(xz)-{1\over h}[z^{-h+1}x\Phi(t^*,x,z)$$ $$+
\sum_2^{h-1}kt_kz^{k-h}\Phi(t^*,x,z)]\exp(xz)+{1\over h}\sum_{h+1}^{\infty}kt_k
\p_{k-h}\Phi(t^*,x,z)\exp(xz).$$ Divided by $\exp(xz)$ this yields the required
equality.

The $\tau$-function is introduced by Sato's formula
$$\hat{w}(t,z)={\tau(t_1-1/z,t_2-1/2z^2,t_3-1/3z^3,...)\over\tau(t_1,t_2,t_3,
...)}.$$ The numerator of this formula will be denoted as $\tilde{\tau}$.

The Virasoro constraint imposed on the $\tau$-function is the equality
\footnote{There are also higher constraints which, as it was shown in \cite{2},
follow from this one.}
$$\sum_{h+1}^{\infty}kt_k\p_{k-h}\tau+{1\over 2}\sum_{k+l=h}klt_kt_l\tau=c\tau.
\eqno{(14)}$$ This equation is a little bit more general than usually written
where $c=0$.

As it was said in the Sect.2, a Baker function of KP is determined up to a
gauge: the multiplication by a constant series $1+\sum_0^{\infty}c_iz^{-i-1}$.
Therefore, a $\tau$-function is also determined up to a gauge: it can have an
arbitrary factor $\exp\sum_1^{\infty}c_kt_k$. However, the string equation is
not invariant under this transformation (see Remark in Sect2); Eq.(14) is not
invariant under gauge transformation.

\proclaim{Proposition} The equation (14) for the $\tau$-function is equivalent
to independence of $\hat{w}$ and $L$ of the variable $\ta$; if, in addition,
$\tau$ does not depend on $t_h$ then  $Q=L^h$ is a differential operator, and
$Q$ and $P=h^{-1}(ML^{-h+1})_+$ satisfy the string equation.\endproclaim

\demo{Proof} First we notice that Eq(14) implies
$$\sum_{h+1}^{\infty}k(t_k-{1\over kz^k})\p_{k-h}\tilde{\tau}+{1\over 2}\sum_{k
+l=h}kl(t_k-{1\over kz^k})(t_l-{1\over lz^l})\tilde{\tau}=c\tilde{\tau}.\eqno{(
15)}$$
Then $$f(t_1,t_2,...;z)\buildrel def\over ={1\over h}\sum_{h+1}^{\infty}k(t_k-
{1\over kz^k})\p_{k-h}
\tilde{\tau}/\tilde{\tau}+{1\over 2h}\sum_{k+l=h}kl(t_k-{1\over kz^k})(t_l-{1
\over lz^l})$$ $$-{1\over h}\sum_{h+1}^{\infty}kt_k\p_{k-h}\tau/\tau-{1\over 2h
}\sum_{k+l=h}klt_kt_l=0.\eqno{(16)}$$ Conversely, if $f(t_1,t_2,...;z)\equiv0$
then $$g(t_1-1/z,t_2-1/2z^2,...)\buildrel def\over =\sum_{h+1}^{\infty}k(t_k-
{1\over kz^k})\p_{k-h}
\tilde{\tau}/\tilde{\tau}+{1\over 2}\sum_{k+l=h}kl(t_k-{1\over kz^k})(t_l-{1
\over lz^l})$$ does not depend on $z$ which yields $\pz g(t_1-1/z,t_2-1/2z^2,..
.)=\sum_1^\infty\p_kgz^{-k-1}=0$. Multiplying by $z^2$ and sending $z
\rightarrow\infty$ we get $\p_1g=0$, multiplying by $z$ and sending $z$ to
infinity again we get $\p_2g=0$ etc, all $\p_jg=0$ and $g=$const which yields
Eqs(15) and (14).
\proclaim{Lemma} $$\pa\hat{w}=f(t;z){\tilde{\tau}\over\tau}.$$ \endproclaim
{\it Proof of the lemma}. Transform the last term of Eq(13) using $\hat{w}=
\tilde{\tau}/\tau$, definition of the function $f(t_1,t_2,...;z)$ and $\pz
\tilde{\tau}=\sum_1^{\infty}z^{-k-1}\p_k\tilde{\tau}$:
$${1\over h}\sum_{k=h+1}^{\infty}kt_k{\tau\p_{k-h}\tilde{\tau}-\tilde{\tau}\p
_{k-h}\tau \over\tau^2}={1\over h}\sum_{k=h+1}^{\infty}k(t_k-{1\over kz^k}){\p_
{k-h}\tilde{\tau}\over\tau}$$ $$-{1\over h}\sum_{k=h+1}^{\infty}kt_k{\p_{k-h}
\tau\over\tau^2}\tilde{\tau}+{1\over h}\sum_{k=h+1}^{\infty}{1\over z^k}{\p_{k
-h}\tilde{\tau}\over\tau}$$ $$=-{1\over 2h}\sum_{k+l=h}kl(t_k-{1\over
kz^k})(t_l-{1\over lz^l}){\tilde{\tau}\over\tau}+{1\over2h}\sum_{k+l=h}klt_kt_l
{\tilde{\tau}\over\tau}$$ $$+{1\over h}\sum_{k=h+1}^{\infty}{1\over z^k}{\p_{k
-h}\tilde{\tau}\over\tau}+f(t;z){\tilde{\tau}\over\tau}
={1\over h}\sum_1^{h-1}kt_kz^{k-h}\hat{w}-{1\over 2h}
\sum_{k+l=h}{1\over z^h}\hat{w}$$ $$+{1\over h}\sum_{k=h+1}^{\infty}{1\over z^k
}{\p_{k-h}\tilde{\tau}\over\tau}+f(t;z){\tilde{\tau}\over\tau}=-{1\over 2h}
(h-1)z^{-h}\hat{w}+{1\over h}\sum_1^{h-1}kt_kz^{k-h}\hat{w}$$ $$+{1\over h}z^
{-h+1}\sum_1^{\infty}z^{-k-1}\p_k\tilde{\tau}\cdot{1\over\tau}+f(t;z){\tilde{
\tau}\over\tau}$$ $$=-{1\over 2h}(h-1)z^{-h}\hat{w}+{1\over h}\sum_1^{h-1}kt_kz
^{k-h}\hat{w}+{1\over h}z^{-h+1}\pz\hat{w}+f(t;z){\tilde{\tau}\over\tau}.$$
Substituting this for the last term of (13) we find that $$\pa\hat{w}=
f(t;z){\tilde{\tau}\over\tau}.$$ Lemma is proved. Now, the equation $\pa\hat{w}
=0$ is equivalent to $f(t;z)=0$ and, therefore, to (14) which proves the
proposition. \enddemo

{\bf References.}
\roster
\item Brezin, E. and Kazakov, V., Phys. Lett. 236 B (1990) 144

Douglas, M. and Shenker, S., Nucl. Phys., B335 (1990) 635

Gross, D. and Migdal, A., Phys. Rev. Lett. 64 (1990) 127

Douglas, M., Phys, Lett. B238 (1990) 176

\item Fukuma, M.,Kawai and H., Nakayama, R., Infinite dimensional Grassmannian
structure of two-dimensional quantum gravity, Commun. Math. Phys., 143 (1992)
371
\item Kac, V. and Schwarz, A., Geometric interpretation of the partition
function of 2D gravity, Phys. Lett. B257 (1991) 329

Schwarz, A., On some mathematical problems of 2D-gravity and $W_h$-gravity,
Modern Pys. Lett. A6 (1991) 611

Schwarz, A., On solutions to the string equation, Preprint Univ.
California, Davis, 1992
\item Dijkgraaf, R., Verlinde, E. and Verlinde, H., Loop equations and
Virasoro constraints in non-perturbatibe two-dimensional quantum gravity,
Nuclear Physics B348 (1991) 435
\item La, HoSeong, Geometry of Virasoro constraints in nonperturbative 2-d
quantum gravity, Commun. Math. Physics 140 (1991) 569

La, HoSeong, Symmetries in non-perturbative 2-d quantum gravity, Modern
Physics Lett. A6 no7 (1991) 573
\item Goeree, J., W-constraints in 2D quantum gravity, Nucl. Phys. B358 (1991)
737
\item Orlov, A. Yu. and Shulman, E.I., Additional symmetries for integrable
equations and conformal algebra representations, Lett. Math. Phys. 12 (1986)
171
\item Dickey, L.A., Soliton equations and integrable systems, World Scientific,
(1991)
\item Date, E., Jimbo, M., Kashiwara, M. and Miwa, T., Transformation groups
for soliton equations, in Jimbo and Miwa (ed) Non-linear integrable systems --
classical theory and quantum theory, Proc. RIMS symposium, Singapore (1983)
\item Gelfand, I. M. and Dickey, L.A., Asymptotics of the resolvent of
Sturm-Liouville's equation and algebra of the KdV equation, Russ. Math.
Surveys, 30 no5 (1975),77

Gelfand, I. M. and Dickey, L.A., Fractional powers of operators and
Hamiltonian systems, Funkt. Anal. Appl., 10 no2 (1976) 75
\item Chen, H. H., Lee, Y. C. and Lin, J. E., On a new hierarchy of symmetries
for the Kadomtsev-Petviashvili equation, Physica D 9D no3 (1983) 439
\item Mulase, M., Category of vector bundles on algebraic curves and infinite
dimensional Grassmannians, Internat. Journal of Math., 1 no3 (1990) 293
\item Segal, G. and Wilson, G., Loop groups and equations of KdV-type,
Publ. Math. IHES 63 (1985) 1-64

\endroster

\enddocument